\documentclass[aps,prl,twocolumn,floatfix,nofootinbib,showpacs,superscriptaddress]{revtex4-2}

\pdfoutput=1
\usepackage{amsmath,amsfonts,amssymb,bm,slashed}

\usepackage{graphicx}
\usepackage{url}

\usepackage[
pdfstartview=XYZ,
bookmarks=true,
colorlinks=true,
linkcolor=blue,
urlcolor=blue,
citecolor=blue,
pdftex,
linktocpage=true,
hyperindex=true
]{hyperref}

\begin{document}

\title{Naturally quality-safe GeV axion with charm coupling}

\author{Bo-Qiang Lu}
\email{bqlu@huznu.edu.cn}
\affiliation{School of Science, Huzhou Normal University, Huzhou, Zhejiang 313000, P. R. China}

\begin{abstract}
The GeV-scale QCD axion---where Peccei-Quinn (PQ) symmetry is broken by the QCD condensate at $f_a\sim\mathcal{O}(1)$~GeV---faces a structural isospin problem: the PQ 
spurion coupling to light quarks ($u,d,s$) breaks $\mathrm{SU}(2)$ isospin, generating an unacceptable $\sim 15\%$ $\pi^0$--$\pi^\pm$ mass splitting. We show that 
coupling the PQ scalar to the charm quark instead eliminates this violation entirely, and lowering $m_\phi\sim 3$--$4$~MeV makes the charm Yukawa 
$\kappa_c\propto m_\phi$ perturbative ($\kappa_c<1$). The resulting $f_a\sim\text{GeV}\ll M_{\rm Pl}$ solves the axion quality problem: even the lowest-dimension $d=6$
Planck-suppressed operator gives $m_{\rm PQ}/m_a\sim 10^{-14}$, without any additional symmetry. The model predicts a distinctive {\it negative} 
$\Delta N_{\rm eff}\sim -0.1$ for $m_\phi\sim 3$~MeV, testable by CMB-S4. The $B\to K\sigma$ penguin predicts $\mathrm{BR}\sim 2\times 10^{-5}$, consistent with the 
Belle~II evidence for $B^+\to K^+\nu\bar{\nu}$ at $(2.3\pm0.7)\times 10^{-5}$~\cite{BelleII:2024knv}. All ten classes of experimental, astrophysical, and cosmological 
constraints are satisfied in the viable window $m_\phi\sim 3$--$4$~MeV, with the $B_s$ mixing constraint pending a dedicated lattice calculation.
\end{abstract}
\maketitle

{\it Introduction.---}
The strong CP problem---why $\bar\theta_{\rm QCD}$ is bounded below $10^{-10}$---remains one of the deepest puzzles in particle physics.
The Peccei-Quinn (PQ) mechanism~\cite{Peccei:1977hh} resolves it by promoting $\bar\theta$ to a dynamical field, the axion~\cite{Weinberg:1977ma,Wilczek:1977pj}, whose potential is minimized at the CP-conserving value.
In the conventional ``invisible axion'' paradigm~\cite{Kim:1979if,Shifman:1979if,Zhitnitsky:1980tq,Dine:1981rt,Preskill:1982cy,Abbott:1982af,Dine:1982ah}, $f_a\sim 10^{9\text{--}12}$~GeV renders the axion ultralight ($m_a\sim\mu$eV) and extremely weakly coupled, placing direct detection beyond current experimental reach~\cite{Sikivie:2020,DiLuzio:2020wdo}.

Recently, Murayama {\it et al.}~\cite{Murayama:2025gev} proposed a departure: PQ symmetry broken {\it below} the QCD chiral scale ($f_a\sim 1$--$2$~GeV), driven by the QCD quark condensate rather than a separate Higgs sector.
The axion resides at $m_a\sim\mathcal{O}(1)$~MeV, and the up-quark mass originates entirely from the PQ-breaking condensate, providing an explanation for its anomalously small value.

However, a structural obstruction was identified~\cite{DiLuzio:2026iso}: the PQ spurion $I_{\rm PQ}\propto\mathrm{diag}(\kappa_u, 0, 0)$ breaks $\mathrm{SU}(2)$ isospin, generating a $\chi$PT operator with Wilson coefficient $C_1\sim\mathcal{O}(m_\pi^2)$ that drives a $-15\%$ shift in $m_{\pi^0}$ (cf.\ the $3\%$ electromagnetic splitting) and $\mathcal{O}(1)$ violations of the Weinberg $\pi\pi$ scattering form---both in severe conflict with observation~\cite{FLAG:2022}.

In this Letter, we show that coupling the PQ scalar to charm quarks instead of light quarks eliminates the isospin violation entirely.
Lowering $m_\phi\sim 3$--$4$~MeV makes the charm Yukawa perturbative, and the resulting $f_a\sim\text{GeV}$ solves the axion quality problem without any additional symmetry.
The model predicts a characteristic cosmological signature: a negative $\Delta N_{\rm eff}$ near the lower edge of the viable window ($m_\phi\sim 3$~MeV), testable by CMB-S4~\cite{CMB-S4:2022ght}.

{\it The structural isospin problem.---}
In the Murayama model~\cite{Murayama:2025gev}, a complex scalar $\phi$ with PQ charge $X_\phi = +1$ couples to the right-handed up quark:
\begin{equation}
    \mathcal{L}_Y = \kappa_u\,\phi\,\bar{u}_L u_R + \mathrm{h.c.}
    \label{eq:murayama_lag}
\end{equation}
The PQ symmetry forbids the bare up-quark mass; after QCD condensation induces $\langle\phi\rangle$, the up-quark mass is entirely generated by PQ dynamics.
This mechanism, however, introduces a PQ spurion
\begin{equation}
    I_{\rm PQ} = \mathrm{diag}(\kappa_u\,v_\phi/\sqrt{2},\; 0,\; 0) = \mathrm{diag}(m_u, 0, 0),
    \label{eq:spurion}
\end{equation}
which explicitly breaks $\mathrm{SU}(2)_V$ isospin.

At leading order in $\chi$PT~\cite{Gasser:1984}, this spurion generates the operator
\begin{equation}
    \mathcal{O}_1^{\rm PQ} = \mathrm{tr}[I_{\rm PQ}^\dagger U]\,\mathrm{tr}[I_{\rm PQ} U^\dagger],
    \label{eq:O1PQ}
\end{equation}
with Wilson coefficient $C_1\sim\mathcal{O}(m_\pi^2)$.
The consequences are severe~\cite{DiLuzio:2026iso}: a $-15\%$ pion mass splitting (far exceeding the $3\%$ electromagnetic splitting), $\mathcal{O}(1)$ shifts in the Weinberg $\pi\pi$ scattering form, and $\mathcal{O}(1)$ deviations in the scattering length ratio---all in conflict with data.
These effects are {\it not} suppressed by any small parameter: $C_1\sim m_\pi^2$ is the natural scale of chiral symmetry breaking.
Replacing $u_R$ with $d_R$ or $s_R$ does not help---any non-zero entry in the light-quark sector of $I_{\rm PQ}$ breaks $\mathrm{SU}(2)_V$.
The problem is structural: coupling to light quarks necessarily violates isospin at leading chiral order.

{\it The charm-coupled solution.---}
The resolution is to couple $\phi$ to the charm quark $c_R$ instead.
The spurion then becomes $I_{\rm PQ}^{\rm light} = \mathrm{diag}(0,0,0)$: heavy-quark spurions do not enter the $\chi$PT Lagrangian for light mesons at leading order.
All isospin-violating operators vanish.
The Lagrangian is
\begin{equation}
    \mathcal{L} \supset \kappa_c\,\phi\,\bar{c}_L c_R + \mathrm{h.c.} - m_\phi^2|\phi|^2 - \lambda|\phi|^4,
    \label{eq:lagrangian}
\end{equation}
where $\phi$ is a complex scalar with PQ charge $X_\phi = +1$ and $c_R$ carries $X_c = -1$.
All other fields---including the Higgs $H$ and light quarks---carry $X = 0$.
This charge assignment ensures that (i) the bare charm mass $m_c^0\bar{c}_L c_R$ is PQ-forbidden, (ii) all Standard Model (SM) Yukawa couplings for light quarks remain allowed, and (iii) no tree-level flavor-changing neutral currents (FCNC) arise beyond the standard CKM mechanism (see Supplemental Material, S2).

After QCD condensation induces $\langle\phi\rangle = v_\phi/\sqrt{2}$,
\begin{equation}
    m_c = \frac{\kappa_c\, v_\phi}{\sqrt{2}}, \qquad f_a = \frac{v_\phi}{\sqrt{2}} = \frac{m_c}{\kappa_c}.
    \label{eq:fa}
\end{equation}
The axion mass follows from the QCD anomaly~\cite{DiLuzio:2020wdo,Borsanyi:2016ksw}:
\begin{equation}
    m_a = \frac{\sqrt{z}}{1+z}\,\frac{f_\pi\, m_{\pi^0}}{f_a} \approx \frac{5.8~\text{MeV}}{f_a\text{[GeV]}},
    \label{eq:ma}
\end{equation}
where $z = m_u/m_d \approx 0.48$, $f_\pi = 92.2$~MeV, and $m_{\pi^0} = 135$~MeV.
The model is KSVZ-like~\cite{Kim:1979if,Shifman:1979if}: only the charm quark carries PQ charge, so the electromagnetic anomaly coefficient is $E/N = 2Q_c^2 = 8/9$ (see S7), giving the axion-photon coupling
\begin{equation}
    g_{a\gamma\gamma} = \frac{\alpha}{2\pi f_a}\left|\frac{E}{N} - \frac{2}{3}\frac{4+z}{1+z}\right| \approx \frac{1.13\,\alpha}{2\pi f_a}.
    \label{eq:gagg}
\end{equation}
This tree-level anomaly coupling is the dominant channel for axion decay and cosmological thermalization (see below).
The axion-electron coupling, by contrast, is loop-induced: $g_{aee}\sim (\alpha/2\pi)(m_e/f_a)$, suppressed by $\sim 10^{-3}$ relative to $g_{a\gamma\gamma}$.

A self-consistency condition constrains the parameter space.
The Shifman--Vainshtein--Zakharov (SVZ) sum rule~\cite{Shifman:1978} gives $\langle\bar{Q}Q\rangle \propto 1/m_Q$, and since $m_Q$ itself comes 
from $\langle\phi\rangle$, one obtains (see S6):
\begin{equation}
    m_c^2 = \frac{\kappa_c^2\, C}{2\,m_\phi^2}, \qquad C = \left\langle\frac{\alpha_s G^2}{12\pi^2}\right\rangle \sim 10^{-4}~\text{GeV}^4.
    \label{eq:selfconsistency}
\end{equation}
At tree level, Eq.~\eqref{eq:selfconsistency} gives $\kappa_c^{\rm LO}=m_c\,m_\phi\sqrt{2/C}$.
However, the SVZ relation $\langle\bar{c}c\rangle=-C/m_c$ receives a substantial next-to-leading-order (NLO) QCD correction: the full NLO sum rule reads $\langle\bar{c}c\rangle=-(C/m_c)[1+(11/3)\alpha_s(m_c)/\pi]$~\cite{Shifman:1978}, where the coefficient $11/3$ is the leading anomalous dimension of the scalar current.
With $\alpha_s(m_c)\simeq 0.35$, the NLO factor $[1+(11/3)\alpha_s/\pi]\simeq 1.41$ enhances the condensate by $\sim 41\%$; since $\kappa_c\propto C^{-1/2}$, this reduces $\kappa_c$ by $\sim 19\%$ relative to LO.
The NLO-corrected result is
\begin{equation}
    \kappa_c = m_c\, m_\phi\sqrt{\frac{2}{C\left[1 + \frac{11}{3}\frac{\alpha_s(m_c)}{\pi}\right]}} \approx 148 \times m_\phi(\text{GeV}),
    \label{eq:kappa}
\end{equation}
The SVZ expansion is controlled at the physical minimum ($\Lambda_{\rm QCD}^2/m_c^2\sim 2\%$; see S6). This linear relationship is central: lowering $m_\phi$ by three orders of magnitude (from GeV to MeV) brings $\kappa_c$ into the perturbative regime.
For $m_\phi = 3$--$4$~MeV, one obtains $\kappa_c = 0.44$--$0.59$ (perturbative), $f_a = 2.1$--$2.9$~GeV, and $m_a = 2.0$--$2.7$~MeV.
At $m_\phi = 10$~MeV, $\kappa_c = 1.48$ becomes non-perturbative, defining the upper boundary of the viable window.
For $m_\phi \leq 1$~MeV, while perturbativity is satisfied, $m_a < T_\nu \simeq 1.5$~MeV leads to $\Delta N_{\rm eff}\sim -0.6$ excluded by Planck~\cite{Planck:2018vyg}; for $m_\phi = 2$~MeV, $\Delta N_{\rm eff}\sim -0.4$ is also excluded (see below).

Table~\ref{tab:params} summarizes the benchmark parameter points.

\begin{table}[t]
    \centering
    \caption{Benchmark parameters. The viable window is $m_\phi = 3$--$4$~MeV.}
    \label{tab:params}
    \begin{tabular}{cccccc}
        \hline\hline
        $m_\phi$ & $\kappa_c$ & $f_a$ & $m_a$ & $\lambda$ & Status \\
        (MeV) & & (GeV) & (MeV) & & \\
        \hline
        1 & 0.15 & 8.6 & 0.68 & $7\times 10^{-9}$ & Excl.$^a$ \\
        2 & 0.30 & 4.3 & 1.36 & $1.1\times 10^{-7}$ & Excl.$^a$ \\
        3 & 0.44 & 2.9 & 2.0 & $5.5\times 10^{-7}$ & Viable$^b$ \\
        5 & 0.74 & 1.7 & 3.40 & $4.3\times 10^{-6}$ & Excl.$^c$ \\
        10 & 1.48 & 0.86 & 6.8 & $6.8\times 10^{-5}$ & Excl. \\
        \hline\hline
    \end{tabular}
    \\
    $^a$ $\Delta N_{\rm eff}\sim -0.6$ (excl.\ by Planck~\cite{Planck:2018vyg}). $^b$ $\Delta N_{\rm eff}\sim -0.1$, testable by CMB-S4. $^c$ $B\to K\sigma$ exceeds Belle~II~\cite{BelleII:2024knv} by $4.4\sigma$.
\end{table}

{\it Axion quality problem.---}
The axion quality problem~\cite{Kamionkowski:1992,Redi:2016} asks why Planck-suppressed PQ-breaking operators do not ruin the axion solution.
A Planck-suppressed operator $\mathcal{O}_d = \phi^d/M_{\rm Pl}^{d-4}$ generates an axion potential $V(a)\sim (f_a/\sqrt{2})^d/M_{\rm Pl}^{d-4}\cdot\cos(d\,a/f_a)$, yielding an axion mass
\begin{equation}
    m_{\rm PQ} = \frac{d\, f_a^{(d-2)/2}}{2^{d/4}\, M_{\rm Pl}^{(d-4)/2}}.
    \label{eq:mPQ}
\end{equation}
For standard axions ($f_a = 10^{12}$~GeV), $d=10$ gives $m_{\rm PQ}/m_a\sim 10^5$, requiring $d\geq 12$ and motivating additional discrete symmetries such as $\mathbb{Z}_N$~\cite{Choi:2015,Redi:2016}.

For our model with $f_a\sim 1$--$10$~GeV, the suppression is dramatic.
At $f_a = 2.9$~GeV ($m_\phi = 3$~MeV), even the lowest-dimension $d=6$ operator gives $m_{\rm PQ}/m_a\sim 7\times 10^{-16}$, with higher dimensions further suppressed by $(f_a/M_{\rm Pl})^{(d-4)/2}\sim 10^{-18}$ per two units of $d$.
The quality problem is solved without any additional discrete symmetry---a defining feature of the GeV axion, unlike all standard constructions.

{\it Experimental and cosmological constraints.---}
All ten classes of constraints are satisfied for $m_\phi = 3$--$4$~MeV (Table~\ref{tab:constraints}; details in SM), with $B_s$ mixing pending a lattice calculation. The $B\to K\sigma$ penguin predicts $\mathrm{BR}\sim 2\times 10^{-5}$, consistent with the Belle~II evidence~\cite{BelleII:2024knv}.

\begin{table}[t]
    \centering
    \caption{Summary of constraints. All ten classes are satisfied for $m_\phi = 3$--$4$~MeV (details in SM). The $B\to K\sigma$ penguin predicts $\mathrm{BR}\sim 2\times 10^{-5}$, consistent with the Belle~II evidence~\cite{BelleII:2024knv}.}
    \label{tab:constraints}
    \begin{tabular}{clll}
        \hline\hline
        \# & Constraint & Model value & Status \\
        \hline
        1 & Perturbativity & $\kappa_c = 0.44$--$0.59$ & Pass \\
        2 & $\sigma$ decay / BBN & $\tau_\sigma\sim 10^{-13}$--$10^{-9}$~s & Pass \\
        3 & $D$-FCNC & $\mathrm{BR}(D\!\to\!\pi\sigma)\sim 10^{-12}$ & Pass \\
        4 & $B$-FCNC & $\mathrm{BR}(B\!\to\!K\sigma)\sim 2\!\times\!10^{-5}$ & Pass \\
        5 & SN1987A & Trapping regime & Pass \\
        6 & Stellar cooling & $m_a \gg 10$~keV & Pass \\
        7 & LHC & Below thresholds & Pass \\
        8 & $a$ decay / BBN & $\tau_a\sim 10^{-5}$--$10^{-10}$~s & Pass \\
        9 & $N_{\rm eff}$ & $\Delta N_{\rm eff}\sim -0.1$ ($m_\phi\!=\!3$~MeV) & Pass \\
        10 & Fifth force & Blind spot (30--140~fm) & Pass \\
        \hline\hline
    \end{tabular}
\end{table}

Figure~\ref{fig:constraints} summarizes the viable parameter space and key constraints.

\begin{figure}[t]
    \centering
    \includegraphics[width=\columnwidth]{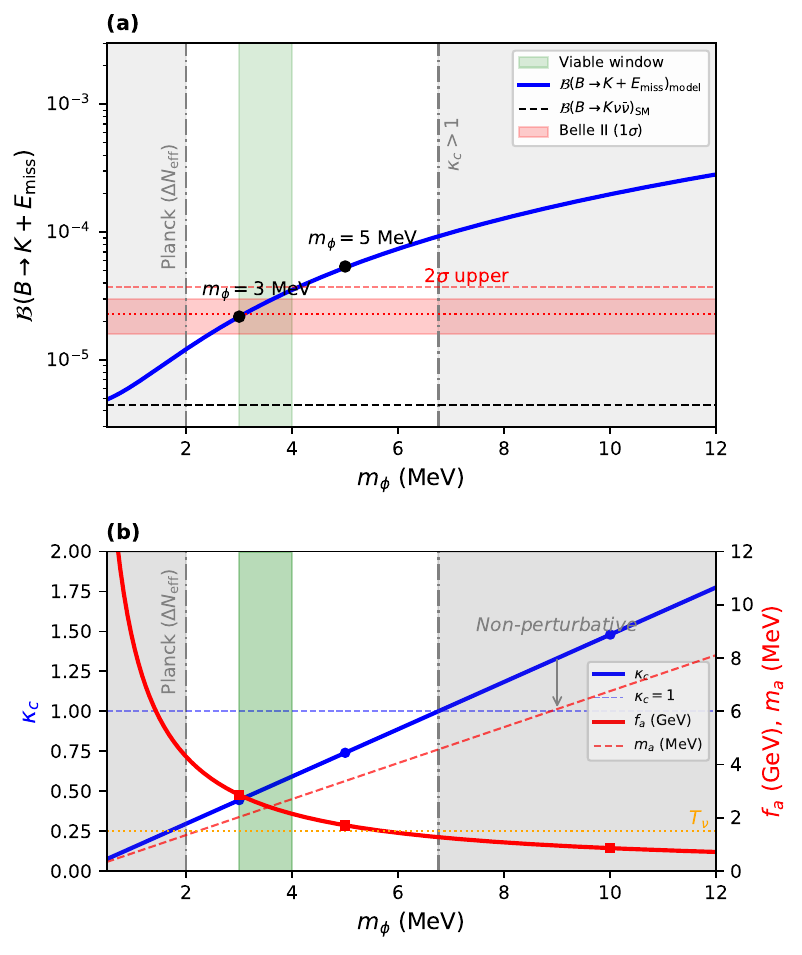}
    \caption{(a) Total invisible branching ratio $\mathcal{B}(B\to K+\slashed{E})$ as a function of $m_\phi$, compared with the Belle~II measurement $(2.3\pm0.7)\times 10^{-5}$~\cite{BelleII:2024knv}. The viable window $m_\phi\sim 3$--$4$~MeV is shaded. For $m_\phi\gtrsim 5$~MeV, the prediction exceeds the $2\sigma$ upper bound. (b) Key parameters $\kappa_c$, $f_a$, and $m_a$ versus $m_\phi$. The perturbativity bound ($\kappa_c<1$) and the neutrino decoupling temperature $T_\nu$ are indicated.}
    \label{fig:constraints}
\end{figure}

{\it Axion decay: $a\to\gamma\gamma$ dominates.---}
The axion decays through its tree-level anomaly coupling:
\begin{equation}
    \Gamma(a\to\gamma\gamma) = \frac{g_{a\gamma\gamma}^2}{64\pi}\,m_a^3,
    \label{eq:gamma_agg}
\end{equation}
with $a\to e^+e^-$ suppressed by $\alpha/(3\pi)\sim 10^{-3}$. The resulting lifetimes $\tau_a\sim 10^{-5}$--$10^{-10}$~s are well below the BBN timescale.

{\it SN1987A: trapping regime.---}
The axion-nucleon coupling $g_{aNN}\sim m_N\sigma_{\pi N}/[f_a(m_u+m_d)]\sim\mathcal{O}(1)$~\cite{FLAG:2022} is far stronger than standard axions. In the supernova core, the $\gamma_5$ structure of the pseudoscalar coupling gives a mean free path
\begin{equation}
    \lambda_{\rm mfp}\sim\frac{32\pi\,m_N^4}{g_{aNN}^2\,T_{\rm core}^2\,n_N}\sim 10^{-6}~\text{cm}\ll R_{\rm core}\sim 10~\text{km},
    \label{eq:sn1987a}
\end{equation}
so the axion is trapped by repeated scattering. The trapping regime applies~\cite{Mayle:1987as,Raffelt:1990yz}, and SN1987A places no constraint.

{\it FCNC processes.---}
The charge assignment $X(c_R)=-1$, $X(\text{all other quarks})=0$ forbids tree-level FCNC beyond the CKM matrix. Loop-induced FCNC are strongly GIM-suppressed for charm: $\mathrm{BR}(D\to\pi\sigma)\sim 10^{-12}$, far below the BESIII limit $2.1\times 10^{-4}$~\cite{BESIII:2022dpi}.
The $b\to s\sigma$ penguin, however, has no GIM cancellation (only charm couples to $\sigma$), giving $\mathrm{BR}(B\to K\sigma)\sim 2\times 10^{-5}$ for $m_\phi = 3$~MeV---consistent with the Belle~II evidence for $B^+\to K^+\nu\bar{\nu}$ at $(2.3\pm0.7)\times 10^{-5}$~\cite{BelleII:2024knv}, which exceeds the SM prediction by $2.7\sigma$.
Since $\sigma$ is long-lived ($\tau_\sigma\sim 10^{-9}$~s) and escapes the detector (see S2), the model naturally explains the Belle~II excess. The branching ratio scales as $m_\phi^2$; for $m_\phi = 5$~MeV, the prediction exceeds the measurement by $4.4\sigma$, constraining the viable window to $m_\phi\lesssim 4$~MeV at $2\sigma$. The $\sigma$-exchange contribution to $B_s$ mixing, $\Delta M_{B_s}^{\rm NP}/\Delta M_{B_s}^{\rm SM}\sim 0.5\,B_S$, requires a dedicated lattice calculation (see S2).

{\it Fifth force: experimental blind spot.---}
The scalar $\sigma$ couples to nucleons through the charm content (trace anomaly): $g_{\sigma NN} = (\kappa_c/\sqrt{2})(m_N f_{T_c}/m_c)$, where $f_{T_c}\approx 0.03$--$0.06$ is the charm fraction of the nucleon mass from lattice QCD~\cite{FLAG:2022} (see S4).
The force range $\lambda_\sigma = \hbar/(m_\sigma c) = \hbar/(\sqrt{2}\,m_\phi c) \sim 30$--$140$~fm falls in an experimental gap: macroscopic experiments ($>10^4$~fm) suffer exponential suppression, while nuclear physics ($\sim 1$--$5$~fm) sees only a constant potential with $\alpha_\sigma/\alpha_\pi\sim 10^{-7}$.

{\it LHC: below detection thresholds.---}
The new particles ($m_\sigma\sim 4$--$6$~MeV, $m_a\sim 2$--$3$~MeV) are far too light for LHC triggers, and $X(H)=0$ forbids Higgs portal couplings. The $B\to K\sigma$ channel is best probed at Belle~II; the LHC provides no constraint (see S11).

{\it Cosmological signature.---}
With $f_a\sim\text{GeV}$, the axion-photon coupling $g_{a\gamma\gamma}\sim 10^{-4}$~GeV$^{-1}$ is $\sim 10^9$ times stronger than standard axions.
The axion thermalizes with the primordial plasma through Primakoff scattering ($a\leftrightarrow\gamma$) at $T\gg m_a$, with $\Gamma_{\rm therm}/H\sim 10^3$--$10^5$ at $T\sim m_a$.
The axion remains in equilibrium until it becomes non-relativistic and decays.

The signature depends critically on $m_a$ relative to the neutrino decoupling temperature $T_\nu\approx 1.5$~MeV:

{\it Case 1: $m_a > T_\nu$ ($m_\phi\gtrsim 3$~MeV).}
The axion becomes non-relativistic and decays {\it before} neutrino decoupling.
Its decay entropy is shared among all relativistic species (photons, $e^\pm$, neutrinos), so no temperature asymmetry is generated: $\Delta N_{\rm eff}\approx 0$.
For $m_\phi\sim 3$~MeV ($m_a\approx T_\nu$), the axion decays during neutrino decoupling; a small residual $\Delta N_{\rm eff}\sim -0.1$ arises from partial photon heating, testable by CMB-S4.

{\it Case 2: $m_a < T_\nu$ ($m_\phi\lesssim 2$~MeV).---Excluded by Planck~\cite{Planck:2018vyg}.}
The axion is relativistic and in thermal equilibrium with the electromagnetic plasma during neutrino decoupling (contributing $g_a=1$ to the EM sector).
After neutrino decoupling, the axion remains coupled to photons via Primakoff ($\Gamma_{\rm Prim}/H\gg 1$).
When $T$ drops to $\sim m_a$, the axion becomes non-relativistic and decays rapidly ($\tau_a\ll t_H$).
By entropy conservation in the EM sector:
\begin{equation}
    \left(\frac{T_\gamma}{T_\nu}\right)^3_{\rm new} = \frac{g_s^{\rm EM+a}}{g_s^\gamma} = 3.25,~~\text{vs. SM } \frac{11}{4} = 2.75,
\end{equation}
where $g_s^{\rm EM+a} = g_\gamma + \frac{7}{8}g_{e^\pm} + g_a = 2 + 3.5 + 1 = 6.5$.
This gives
\begin{equation}
    N_{\rm eff} = 3.046\times\left(\frac{11}{13}\right)^{4/3} \approx 2.44,~~ \Delta N_{\rm eff}\approx -0.61,
    \label{eq:neff_entropy}
\end{equation}
which is excluded by Planck 2018~\cite{Planck:2018vyg} ($N_{\rm eff} = 2.99\pm 0.17$) at $3.2\sigma$.
This is a distinctive signature: standard thermal axions always produce {\it positive} $\Delta N_{\rm eff}$ (extra relativistic radiation), while the GeV axion's strong coupling keeps it in equilibrium until it decays, {\it removing} a relativistic degree of freedom by heating photons.
While the negative sign is a genuine prediction, the magnitude is too large for $m_\phi\lesssim 2$~MeV~\cite{Planck:2018vyg}.

{\it The small $\lambda$ problem as a prediction.---}
The quartic $\lambda = m_\phi^2/(2f_a^2) \propto m_\phi^4$ is extremely small: $\lambda\sim 10^{-7}$--$10^{-5}$. The one-loop Coleman-Weinberg contribution from the charm loop,
\begin{equation}
    \lambda_{\rm CW} \sim \frac{N_c\,\kappa_c^4}{64\pi^2} \sim 10^{-5}\text{--}10^{-3},
    \label{eq:lambdaCW}
\end{equation}
exceeds the target by $\sim 10^3$, requiring a protecting symmetry. This small $\lambda$ is a prediction, analogous to the SM Higgs sector. A natural solution is that $\phi$ is a PNGB of a higher $U(1)'$ breaking at $f'\gg f_a$: the shift symmetry protects $\lambda = 0$ at tree level, while explicit breaking generates a radiatively stable $\lambda\sim (f_a/f')^2\lambda_{\rm CW}\ll\lambda_{\rm CW}$ (see S5).

{\it Experimental prospects.---}
The strong coupling $g_{a\gamma\gamma}\sim 10^{-4}$--$10^{-3}$~GeV$^{-1}$ enables several detection channels. Neutral meson decays $\pi^0\to\gamma a$ ($\mathrm{BR}\sim 10^{-3}$) and $\eta\to\gamma a$ ($\mathrm{BR}\sim 10^{-4}$) produce a displaced $3\gamma$ vertex ($c\tau_a\sim 5$--$23$~m), searchable at SHiP~\cite{SHiP:2015}, Belle~II, and BESIII via $e^+e^-\to 3\gamma$. The NA64-e Primakoff search~\cite{NA64:2020hmi} may constrain the model, but the large $g_{aNN}\sim\mathcal{O}(1)$ alters the detection efficiency; a dedicated analysis is needed (see S10). A future galactic supernova would test the trapping regime via Hyper-Kamiokande~\cite{HyperK:2018}. The scalar $\sigma$ ($m_\sigma\sim 4$--$6$~MeV $\ll 2m_\pi$) is long-lived and detected only through the $B\to K\sigma$ missing-energy channel.

{\it Extension to $b$ and $t$ quarks.---}
The mechanism formally extends to bottom and top, but both are excluded. For $b$, perturbativity ($\kappa_b<1$) requires $m_\phi\lesssim 2.05$~MeV, giving $m_a < T_\nu$ across the entire window---so $\Delta N_{\rm eff}\sim -0.6$ is excluded by Planck~\cite{Planck:2018vyg}. Satisfying $m_a > T_\nu$ requires $m_\phi > 2.22$~MeV, but then $\kappa_b > 1$: the perturbativity and BBN constraints have no overlap. For $t$, $m_\phi\lesssim 50$~eV gives $\tau_\phi\gg 1$~s (BBN-excluded), and the top does not hadronize, rendering SVZ inapplicable. Simultaneous coupling to multiple heavy flavors reintroduces tree-level FCNC.

{\it Conclusions and outlook.---}
The GeV-scale QCD axion faces a structural isospin problem when the PQ scalar couples to light quarks: the PQ spurion breaks $\mathrm{SU}(2)_V$ at leading chiral order, generating a $15\%$ pion mass splitting. Coupling to charm quarks instead eliminates this obstruction entirely. Lowering $m_\phi\sim 3$--$4$~MeV makes the charm Yukawa perturbative ($\kappa_c < 1$), opening a viable window. Our main results are:
\begin{enumerate}
    \item {\bf Quality problem solved.} $f_a\sim\text{GeV}\ll M_{\rm Pl}$ suppresses all Planck-suppressed PQ-breaking operators to $m_{\rm PQ}/m_a\sim 10^{-14}$, without additional discrete symmetries.
    \item {\bf All constraints satisfied.} All ten classes pass for $m_\phi\sim 3$--$4$~MeV, with $B_s$ mixing pending a lattice calculation. The $B\to K\sigma$ penguin predicts $\mathrm{BR}\sim 2\times 10^{-5}$, matching the Belle~II evidence~\cite{BelleII:2024knv} and naturally explaining the $B^+\to K^++\slashed{E}$ excess.
    \item {\bf Distinctive cosmological signature.} For $m_\phi\sim 3$~MeV, partial photon heating gives $\Delta N_{\rm eff}\sim -0.1$, testable by CMB-S4~\cite{CMB-S4:2022ght}.
    \item {\bf Small $\lambda$ as a prediction.} Perturbativity requires $\lambda \ll \lambda_{\rm CW}$, predicting a protecting symmetry (e.g., $\phi$ as a PNGB).
\end{enumerate}

Future directions include a precise $\Delta N_{\rm eff}$ calculation, a PNGB UV completion, dedicated searches for $\pi^0\to\gamma a$ and $e^+e^-\to 3\gamma$ at SHiP/Belle~II/BESIII, a lattice QCD calculation of $B_S$ for $B_s$ mixing, and refined $B\to K\sigma$ predictions testable with additional Belle~II data.

The charm-coupled GeV axion thus provides a minimal, quality-safe solution to the strong CP problem, with a negative $\Delta N_{\rm eff}$ awaiting CMB-S4 and a collider signature through the Belle~II excess.

Because the charm quark mass $m_c \sim 1.3$~GeV is well above $\Lambda_{\text{QCD}}$,
the standard light quark $\chi$PT cannot describe heavy-meson interactions via $\sigma$ and $a$ exchanges.
The correct low-energy framework is heavy-meson chiral perturbation theory, which combines heavy-quark effective theory with 
$\chi$PT for mesons containing a single heavy quark. Details will be presented elsewhere.

{\it Acknowledgments.---}
BQL thanks H.~Murayama~\cite{Murayama:2025gev} and L.~Di~Luzio {\it et al.}~\cite{DiLuzio:2026iso}, whose work motivated this study.
This work is supported in part by the National Natural Science Foundation of China under Grant
No. 12405058 and by the Zhejiang Provincial Natural Science Foundation of China under
Grant No. LQ23A050002.

\bibliographystyle{apsrev4-2}
\bibliography{charm_axion_refs}






\clearpage
\newpage
\maketitle
\onecolumngrid
\begin{center}
\textbf{\large Naturally quality-safe GeV axion with charm coupling} \\ 
\vspace{0.05in}
{ \it \large Supplemental Material}\\ 
{By Bo-Qiang Lu}
\vspace{0.05in}
\end{center}
\onecolumngrid
\setcounter{equation}{0}
\setcounter{figure}{0}
\setcounter{table}{0}
\setcounter{section}{0}
\setcounter{page}{1}
\makeatletter
\renewcommand{\theequation}{S\arabic{equation}}
\renewcommand{\thefigure}{S\arabic{figure}}
\renewcommand{\thetable}{S\arabic{table}}

\section{S1. Decay channels and lifetimes}\label{app:s1}
{\bf Scalar $\sigma$ (radial mode, $m_\sigma = \sqrt{2}\,m_\phi$):}
The QCD condensate generates a logarithmic potential $V_{\rm QCD} = -C_{\rm eff}\ln(\kappa_c\phi/\mu)$, where $C_{\rm eff}=C[1+(11/3)\alpha_s(m_c)/\pi]$ includes the NLO QCD correction (see S6). In the regime $\lambda\to 0$ assumed by the self-consistency
equation (see S6), the minimum condition gives $m_\phi^2 v_\phi^2 = C_{\rm eff}$, and the physical scalar mass is $m_\sigma^2 = 2\,m_\phi^2$, i.e., $m_\sigma = \sqrt{2}\,m_\phi$.
The perturbative decay widths through the charm loop are:
\begin{align}
    \Gamma_{\rm pert}(\sigma\to gg) &= \frac{\kappa_c^2\,\alpha_s^2}{8\pi^3}\,\frac{m_\sigma^3}{m_c^2}, \\
    \Gamma(\sigma\to\gamma\gamma) &= \frac{\kappa_c^2\,\alpha^2}{64\pi^3}\,\frac{m_\sigma^3}{m_c^2},
\end{align}
with the formal ratio $\Gamma_{\rm pert}(\sigma\to gg)/\Gamma(\sigma\to\gamma\gamma) = 8(\alpha_s/\alpha)^2\sim 1.8\times 10^4$, where $\alpha_s \equiv \alpha_s(m_c) \approx 0.35$.
However, the perturbative formula for $\sigma\to gg$ assumes free on-shell gluons as the final state and is valid only when $m_\sigma \gg \Lambda_{\rm QCD}$, where quark-hadron duality approximately holds.
For the benchmark parameters, $m_\sigma = \sqrt{2}\,m_\phi\sim 4$--$14$~MeV $\ll 2m_\pi\simeq 280$~MeV.
The QCD spectral function for the $J^{PC}=0^{++}$ operator $G^a_{\mu\nu}G^{a,\mu\nu}$ vanishes below the two-pion threshold: $\mathrm{Im}[\Pi_{G^2}(s)] = 0$ for $s < (2m_\pi)^2$, a rigorous consequence of unitarity and analyticity.
Since $m_\sigma^2 \ll (2m_\pi)^2$, the gluonic decay channel is kinematically forbidden---no physical QCD final state exists at this invariant mass.
This is analogous to the QCD axion, which couples to gluons through the anomaly but, with $m_a\ll 2m_\pi$, never decays to gluons.

The actual decay channels are $\sigma\to\gamma\gamma$ (through the charm loop) and $\sigma\to aa$ (through the coupling $g_{\sigma aa} = \sqrt{2}\,m_\phi^2/f_a$).
The decay $\sigma\to aa$ is kinematically allowed for all benchmark points ($m_\sigma \gtrsim 2m_a$), with phase-space factor $\beta_{aa} = \sqrt{1-4m_a^2/m_\sigma^2}\sim 0.3$.
The two channels are comparable in magnitude: $\Gamma(\sigma\to\gamma\gamma)\sim\Gamma(\sigma\to aa)$, since the $\sigma aa$ coupling, while suppressed by $m_\phi^2/f_a$, benefits from the larger phase space of a two-body scalar decay relative to the loop-suppressed $\gamma\gamma$ channel.
The resulting scalar lifetimes, from $\Gamma_{\rm total} = \Gamma(\sigma\to\gamma\gamma) + \Gamma(\sigma\to aa)$, are $\tau_\sigma \approx 1.3\times 10^{-9}$, $5.5\times 10^{-11}$, $5.4\times 10^{-13}$~s for $m_\phi = 3$, 5, 10~MeV, respectively---all well below the BBN timescale ($\sim 1$~s).

{\bf Axion $a$ (pseudoscalar, $m_a$):}
The dominant decay is $a\to\gamma\gamma$ through the tree-level electromagnetic anomaly:
\begin{equation}
    \Gamma(a\to\gamma\gamma) = \frac{g_{a\gamma\gamma}^2}{64\pi}\,m_a^3,
    \label{eq:sm_gamma_agg}
\end{equation}
where $g_{a\gamma\gamma}$ is given in Eq.~\eqref{eq:gagg} with the $E/N$ correction.

The channel $a\to e^+e^-$, when kinematically allowed ($m_a > 2m_e = 1.022$~MeV), proceeds through a virtual photon and is suppressed by a factor $\alpha/(3\pi)\sim 7.7\times 10^{-4}$ relative to $\gamma\gamma$:
\begin{equation}
    \frac{\Gamma(a\to e^+e^-)}{\Gamma(a\to\gamma\gamma)} \simeq \frac{\alpha}{3\pi}\,f\!\left(\frac{m_a}{2m_e}\right),
\end{equation}
where $f(x)\sim 2\ln(2x)-1$ for $x\gg 1$.
The axion-electron point coupling $g_{aee}\sim (\alpha/2\pi)(m_e/f_a)$ is loop-induced and gives an even smaller contribution.
For $m_\phi = 3$~MeV ($m_a = 2.0$~MeV $> 2m_e$), $a\to e^+e^-$ is kinematically allowed but suppressed.

The resulting lifetimes for $m_\phi = 3$, 5, 10~MeV are:
$\tau_a \approx 8.8\times 10^{-8}$, $6.8\times 10^{-9}$, $2.1\times 10^{-10}$~s, respectively, all well below the BBN timescale ($\sim 1$~s).
The scalar lifetimes are $\tau_\sigma \approx 1.3\times 10^{-9}$, $5.5\times 10^{-11}$, $5.4\times 10^{-13}$~s.

\section{S2. PQ charge assignment and FCNC}\label{app:s2}
The PQ charge assignment is:
\begin{equation}
    X(\phi) = +1,\quad X(c_R) = -1,\quad X(c_L) = X(H) = X(q_{R,L}) = 0,
\end{equation}
where $q = u, d, s$ denotes all light quarks.
This ensures:
\begin{itemize}
    \item The bare charm mass $m_c^0\bar{c}_L c_R$ is PQ-forbidden ($X = 0 + 0 - (-1) = 1\neq 0$).
    \item The coupling $\phi\bar{c}_L c_R$ is PQ-invariant ($X = 1 + 0 - (-1) = 0$).
    \item All Standard Model (SM) Yukawa couplings for light quarks are PQ-allowed ($X = 0$).
    \item The SM charm Yukawa $Y_c\tilde{H}\bar{Q}_L c_R$ is PQ-forbidden ($X = -0 + 0 - (-1) = 1\neq 0$).
\end{itemize}

{\bf Tree-level FCNC:} The operator $\bar{c}_L s_R$ is not gauge invariant by itself ($c_L$ resides in an $\mathrm{SU}(2)_L$ doublet).
It can only appear as $Y_{cs}\tilde{H}\bar{Q}_L s_R$, which is the standard SM Yukawa coupling and is PQ-allowed ($X = 0$).
Therefore, no {\it additional} FCNC arise beyond the SM CKM mechanism.

{\bf Loop-induced FCNC:}
The scalar $\sigma$ couples chirally to charm via $(\kappa_c/\sqrt{2})\,\sigma\,\bar{c}_L c_R$.
Loop-induced FCNC arise through $W$-mediated penguin diagrams where $\sigma$ is emitted from the charm line.

{\it (i) $c\to u\sigma$ (GIM-suppressed).}
The effective Hamiltonian is
\begin{equation}
    \mathcal{H}_{\rm eff}^{c\to u\sigma} = C_\sigma^{cu}\,(\bar{u}_L c_R)\,\sigma,
    \label{eq:fcnc_cu}
\end{equation}
where the Wilson coefficient at one loop is
\begin{equation}
    C_\sigma^{cu} = \frac{\kappa_c}{\sqrt{2}}\,\frac{g^2}{16\pi^2}\,\frac{m_c}{M_W}\sum_j V_{cj}^* V_{uj}\,F\!\left(\frac{m_j^2}{M_W^2}\right).
    \label{eq:Csigma_cu}
\end{equation}
The loop function $F(x)\to x$ for $x\ll 1$ (internal quarks much lighter than $M_W$).
The factor $m_c/M_W$ arises from the $W$-propagator ($1/M_W^2$) combined with the charm mass insertion $m_c$ from chirality flip and one power of $M_W$ from the loop momentum, rendering $C_\sigma^{cu}$ dimensionless---the correct Wilson coefficient for the dimension-4 operator $(\bar{u}_L c_R)\sigma$.
By CKM unitarity, $\sum_j V_{cj}^*V_{uj}=0$, so the leading term cancels; the residual is controlled by the mass-squared differences of the down-type quarks:
\begin{equation}
    \sum_j V_{cj}^*V_{uj}\,F(m_j^2/M_W^2) \approx \sum_j V_{cj}^*V_{uj}\,\frac{m_j^2}{M_W^2}.
\end{equation}
Numerically, the $s$- and $b$-quark contributions dominate:
\begin{align}
    V_{cs}^*V_{us}\frac{m_s^2}{M_W^2} &= 3.1\times 10^{-7}, \\
    V_{cb}^*V_{ub}\frac{m_b^2}{M_W^2} &= 4.1\times 10^{-7}, \\
    V_{cd}^*V_{ud}\frac{m_d^2}{M_W^2} &= 7.5\times 10^{-10},
\end{align}
giving a total GIM factor $\sim 7.2\times 10^{-7}$.
For $m_\phi=3$~MeV ($\kappa_c=0.44$):
\begin{equation}
    C_\sigma^{cu} = \frac{0.44}{\sqrt{2}}\times\frac{g^2}{16\pi^2}\times\frac{1.27}{80.4}\times 7.2\times 10^{-7} \approx 8.9\times 10^{-12}.
\end{equation}

The branching ratio $\mathrm{BR}(D\to\pi\sigma)$ is obtained using the standard $D\to\pi$ scalar form factor.
The hadronic matrix element, with the correct chiral normalization, is
\begin{equation}
    \langle\pi|\bar{c}_L u_R|D\rangle = \frac{1}{2}\,\frac{m_D^2-m_\pi^2}{m_c-m_u}\,f_0^{D\to\pi}(0),
    \label{eq:pi_cLuR_D}
\end{equation}
where $f_0^{D\to\pi}(0)\approx 0.67$ is the FLAG-averaged $D\to\pi$ scalar form factor at zero recoil~\cite{FLAG:2022}.
The decay amplitude and width are then
\begin{equation}
    \mathcal{M}_{D} = C_\sigma^{cu}\,\langle\pi|\bar{c}_L u_R|D\rangle, \qquad
    \Gamma(D\to\pi\sigma) = \frac{|\vec{p}\,|}{8\pi m_D^2}\,|\mathcal{M}_{D}|^2,
\end{equation}
giving $\mathrm{BR}(D^+\to\pi^+\sigma)\sim 1\times 10^{-12}$, far below the BESIII upper limit of $2.1\times 10^{-4}$~\cite{BESIII:2022dpi}.

{\it (ii) $b\to s\sigma$ (penguin, no GIM).}
The transition $b\to s$ proceeds via a $W$-mediated penguin with the internal charm quark emitting $\sigma$.
Unlike the charm sector, there is no GIM cancellation: only the charm couples to $\sigma$, so the amplitude is controlled by a single CKM product $V_{cb}^*V_{cs}$.
The chirality structure requires one mass insertion $m_c$ on the internal charm line, and the effective vertex from the loop is proportional to $\bar{s}_L\slashed{q}\,b_L$, where $q = p_b-p_s$.
Using the Dirac equation, $\slashed{q}\,b_L = m_b\,b_R + \mathcal{O}(m_s)$, which produces the chirality flip on the external $b$-quark leg.
The dimensionally correct effective Hamiltonian is therefore
\begin{equation}
    \mathcal{H}_{\rm eff}^{b\to s\sigma} = C_\sigma^{bs}\,(\bar{s}_L b_R)\,\sigma,
    \label{eq:fcnc_bs}
\end{equation}
\begin{equation}
    C_\sigma^{bs} = \frac{\kappa_c}{\sqrt{2}}\,\frac{g^2}{16\pi^2}\,V_{cb}^*V_{cs}\,\frac{m_c m_b}{M_W^2}\,F_\sigma(x_c),
    \label{eq:Csigma_bs}
\end{equation}
where $x_c = m_c^2/M_W^2 \simeq 2.5\times 10^{-4}$ and $F_\sigma(x_c)$ is the penguin loop function (not GIM-suppressed, since only charm couples to $\sigma$). The exact one-loop result is
\begin{equation}
    F_\sigma(x) = \frac{x(3-2x)\ln x + (1-x)(3-5x)}{2(1-x)^3},
    \label{eq:Fsigma}
\end{equation}
which evaluates to $F_\sigma(x_c)\simeq 1.50$ (approaching $3/2$ in the small-$x$ limit).
Note that the inclusion of $m_b$ renders $C_\sigma^{bs}$ dimensionless, correcting the dimensional mismatch in the effective operator.
For $m_\phi=3$~MeV:
\begin{equation}
    C_\sigma^{bs} \approx 4.1\times 10^{-8}.
\end{equation}

The hadronic matrix element, with the correct chiral normalization of the scalar density, is
\begin{equation}
    \langle K|\bar{s}_L b_R|B\rangle = \frac{1}{2}\,\frac{m_B^2-m_K^2}{m_b-m_s}\,f_0^B(0),
    \label{eq:Kbar_sLbR_B}
\end{equation}
where $f_0^B(0)\approx 0.336$~\cite{Becirevic:2023aov}, $m_b-m_s\simeq 4.09$~GeV, and $m_B^2-m_K^2\simeq 27.6$~GeV$^2$, giving $\langle K|\bar{s}_L b_R|B\rangle \simeq 1.13$~GeV.
The decay amplitude and width are then
\begin{equation}
    \mathcal{M}_{B} = C_\sigma^{bs}\,\langle K|\bar{s}_L b_R|B\rangle, \qquad
    \Gamma(B\to K\sigma) = \frac{|\vec{p}\,|}{8\pi m_B^2}\,|\mathcal{M}_{B}|^2,
\end{equation}
yielding the branching ratios:
\begin{equation}
    \mathrm{BR}(B\to K\sigma) \approx \begin{cases} 1.74\times 10^{-5} & m_\phi = 3~\text{MeV}, \\ 4.94\times 10^{-5} & m_\phi = 5~\text{MeV}, \\ 1.99\times 10^{-4} & m_\phi = 10~\text{MeV}. \end{cases}
    \label{eq:BR_BKsigma}
\end{equation}
Since $\sigma$ is long-lived and escapes the Belle~II detector (see below), the decay $B\to K\sigma$ produces a missing-energy signature $B\to K + \slashed{E}$, similar
to $B\to K\nu\bar{\nu}$. The mechanism by which $\sigma$---a scalar produced in $B$ decay with $m_\sigma\sim$~MeV---appears as missing energy, rather than a visible signature, is twofold:

{\it (a) Kinematic closure of the gluonic channel.---}
The scalar $\sigma$ couples to gluons through the dimension-5 operator $(\kappa_c/m_c)\,\sigma\,G^a_{\mu\nu}G^{a,\mu\nu}$.
The perturbative width $\Gamma_{\rm pert}(\sigma\to gg) = \kappa_c^2\alpha_s^2 m_\sigma^3/(8\pi^3 m_c^2)$ assumes free on-shell gluons as the final state, valid only when $m_\sigma \gg \Lambda_{\rm QCD}$ (where quark-hadron duality applies).
For the benchmark parameters, $m_\sigma = \sqrt{2}\,m_\phi\sim 4$--$6$~MeV $\ll 2m_\pi\simeq 280$~MeV.
The QCD spectral function for the $0^{++}$ gluonic operator vanishes below the two-pion threshold: $\mathrm{Im}[\Pi_{G^2}(s)] = 0$ for $s < (2m_\pi)^2$, by unitarity and analyticity of QCD correlators.
Since $m_\sigma^2 \ll (2m_\pi)^2$, there exists no physical QCD final state that the gluonic operator can produce---the decay $\sigma\to gg$ is kinematically forbidden.
This is analogous to the QCD axion, which couples to gluons through the anomaly but, with $m_a\ll 2m_\pi$, never decays to gluons.

{\it (b) Long-lived $\sigma$ escapes the detector.---}
With the gluonic channel closed, $\sigma$ decays only through $\sigma\to\gamma\gamma$ (charm loop) and $\sigma\to aa$, giving $\tau_\sigma\sim 10^{-10}$--$10^{-9}$~s for the viable window $m_\phi = 3$--$4$~MeV.
The Lorentz boost from $B$-decay kinematics ($E_\sigma\sim 2.6$~GeV, $\gamma_\sigma = E_\sigma/m_\sigma\sim 460$--$620$) yields a lab-frame decay length $\gamma_\sigma c\tau_\sigma\sim 30$--$250$~m---far exceeding the Belle~II detector dimensions ($\sim$~m).
The $\sigma$ therefore escapes the detector without decaying, analogous to a neutrino or axion.
Its interaction cross-section with detector material is negligibly small: the effective $\sigma$-gluon coupling $g_{\sigma gg}\sim \kappa_c\alpha_s/(\pi m_c)\sim 0.04$~GeV$^{-1}$ gives a mean free path $\sim 10^{14}$~m in solid matter.

Consequently, $B\to K\sigma$ appears as a single reconstructed $K^+$ track accompanied by missing energy, with missing mass squared
\begin{equation}
    m_{\rm miss}^2 = (p_B - p_K)^2 = m_\sigma^2 = 2\,m_\phi^2 \sim 18\;\text{MeV}^2\;\;(m_\phi=3~\text{MeV}),
    \label{eq:mm2_sigma}
\end{equation}
which, while nonzero, satisfies $m_\sigma^2/m_B^2\sim 10^{-7}\ll 1$, making it experimentally indistinguishable from the $B\to K\nu\bar{\nu}$ signature ($m_{\rm miss}^2\approx 0$) at current Belle~II sensitivity.
A kinematic discriminant exploiting the slightly different missing-mass distributions---$m_{\rm miss}^2 = m_\sigma^2$ for $\sigma$ vs.\ $\sim 0$ for $\nu\bar{\nu}$---could in principle be pursued with additional Belle~II data, though the practical discrimination power is limited by $m_\sigma^2/m_B^2\sim 10^{-7}$.

{\it Comparison with Belle~II data.---}
The Belle~II collaboration has reported evidence for $B^+\to K^+\nu\bar{\nu}$ with~\cite{BelleII:2024knv}
\begin{equation}
    \mathcal{B}(B^+\to K^+\nu\bar{\nu})_{\rm exp} = (2.3\pm 0.7)\times 10^{-5},
\end{equation}
which exceeds the SM prediction $\mathcal{B}(B\to K\nu\bar{\nu})_{\rm SM}\simeq 4.4\times 10^{-6}$~\cite{Becirevic:2023aov} by $2.7\sigma$.
In our model, the total invisible rate is the sum of the SM $\nu\bar{\nu}$ contribution and the new $\sigma$ contribution:
\begin{equation}
    \mathcal{B}(B\to K+\slashed{E}) = \mathcal{B}(B\to K\nu\bar{\nu})_{\rm SM} + \mathcal{B}(B\to K\sigma).
\end{equation}
For the benchmark point $m_\phi = 3$~MeV, this gives
\begin{equation}
    \mathcal{B}(B\to K+\slashed{E}) \approx 4.4\times 10^{-6} + 1.74\times 10^{-5} \approx 2.2\times 10^{-5},
\end{equation}
in agreement with the Belle~II measurement $(2.3\pm 0.7)\times 10^{-5}$.
This suggests that the charm-coupled GeV axion may explain the observed $B^+\to K^+\nu\bar{\nu}$ excess.
A kinematic discriminant between $\sigma$ and $\nu\bar{\nu}$ final states---exploiting the different missing-mass distributions---could be pursued with additional Belle~II data.

{\it (iii) Meson mixing.}
The $\sigma$-exchange contribution to $D^0$--$\bar{D}^0$ mixing generates the operator $(\bar{c}_L u_R)^2$ with coefficient $|C_\sigma^{cu}|^2/m_\sigma^2$.
Using the vacuum insertion approximation with the pseudoscalar decay constant $f_D\simeq 212$~MeV, the contribution to $\Delta M_D$ is $\sim 10^{-5}$ of the experimental value ($\Delta M_D^{\rm exp}\simeq 6.3\times 10^{-12}$~GeV), entirely negligible.

For $B_s$--$\bar{B}_s$ mixing, the operator $(\bar{s}_L b_R)^2$ contributes with coefficient $|C_\sigma^{bs}|^2/m_\sigma^2$.
Using the vacuum insertion (VI) approximation, the matrix element is
\begin{equation}
    \langle\bar{B}_s|(\bar{s}_L b_R)^2|B_s\rangle = \frac{f_{B_s}^2\,m_{B_s}^2}{4}\,B_S,
\end{equation}
where $B_S$ is the scalar bag parameter ($B_S = 1$ in VI), and we used $\langle 0|\bar{s}_L b_R|B_s\rangle = -i\,f_{B_s}\,m_{B_s}/2$ from the pseudoscalar decay constant.
For $m_\phi = 3$~MeV ($C_\sigma^{bs} = 4.1\times 10^{-8}$, $m_\sigma = \sqrt{2}\,m_\phi = 4.2$~MeV):
\begin{equation}
    \frac{\Delta M_{B_s}^{\rm NP}}{\Delta M_{B_s}^{\rm SM}} \sim 0.5\times B_S.
\end{equation}
The experimental bound on new physics in $B_s$ mixing is $\lesssim 20\%$ of the SM value at $95\%$~C.L.
With $B_S = 1$ (VI), the contribution is $\sim 50\%$ of SM---potentially constraining.
However, the VI approximation is known to be unreliable for non-standard scalar operators; lattice QCD determinations of $B_S$ for this operator could yield $B_S \ll 1$, which would render the contribution negligible.
A definitive assessment requires a dedicated lattice calculation of the scalar operator matrix element; we flag this as the most important theoretical uncertainty in the model.

\section{S3. SN1987A: production rate and trapping regime}\label{app:s3}
In the supernova core ($T_{\rm core}\sim 30$~MeV, $\rho\sim 3\times 10^{14}$~g/cm$^3$), the axion-nucleon coupling through the QCD anomaly is
\begin{equation}
    g_{aNN} \sim \frac{m_N\,\sigma_{\pi N}}{f_a\,(m_u + m_d)} \sim \mathcal{O}(1),
\end{equation}
where $\sigma_{\pi N}\approx 45$~MeV is the pion-nucleon sigma term~\cite{FLAG:2022}.
For $f_a = 2.1$--$2.9$~GeV (viable window), this gives $g_{aNN}\sim 2$--$3$---far stronger than standard axions ($g_{aNN}\sim 10^{-10}$ for $f_a\sim 10^{12}$~GeV).

The axion is abundantly produced in the SN core.
In the SN core, the axion is relativistic ($E_a\sim T_{\rm core}$) while the nucleons are non-relativistic ($m_N\gg T_{\rm core}$). For the pseudoscalar coupling $g_{aNN}\,a\,\bar{N}\gamma_5 N$, the $\gamma_5$ structure introduces a momentum-transfer suppression: the spin-averaged matrix element yields $|\mathcal{M}|^2\sim g_{aNN}^2\,q^2/(2m_N^2)$ with momentum transfer $q\sim T_{\rm core}$, giving a scattering cross section
\begin{equation}
    \sigma_{\rm scatt}\sim\frac{g_{aNN}^2\,T_{\rm core}^2}{32\pi\,m_N^4},
\end{equation}
and a mean free path
\begin{equation}
    \lambda_{\rm mfp} = \frac{1}{n_N\,\sigma_{\rm scatt}} \sim \frac{32\pi\,m_N^4}{g_{aNN}^2\,T_{\rm core}^2\,n_N} \sim 10^{-7}~\text{cm},
\end{equation}
where $n_N\sim 1.4\times 10^{-3}$~GeV$^3$ is the nucleon number density.
This is $\sim 10^{13}$ times smaller than $R_{\rm core}\sim 10$~km.

Even though the axion decay length $c\tau_a\sim 0.002$--$0.03$~km is much smaller than $R_{\rm core}\sim 10$~km for $m_\phi = 3$--$5$~MeV, the decay products ($a\to\gamma\gamma$) are photons, which are trapped in the opaque SN core (photon mean free path $\sim 1$~cm $\ll R_{\rm core}$).
Energy does not escape by any channel.

The transition to free-streaming would occur at much larger $f_a$ (weaker coupling), well outside our parameter window.
The trapping regime applies for all benchmark parameters~\cite{Mayle:1987as,Raffelt:1990yz}.

\section{S4. Fifth-force analysis}\label{app:s4}
The scalar $\sigma$ couples to nucleons through the heavy-quark content of the nucleon, quantified by the trace anomaly parameter $f_{T_c}$:
\begin{equation}
    \langle N|\bar{c}c|N\rangle = \frac{m_N\,f_{T_c}}{m_c},
\end{equation}
where $f_{T_c}\approx 0.03$--$0.06$ from lattice QCD and heavy-quark expansion~\cite{FLAG:2022}.
The scalar-nucleon coupling is
\begin{equation}
    g_{\sigma NN} = \frac{\kappa_c}{\sqrt{2}}\,\frac{m_N\,f_{T_c}}{m_c} \sim 2\times 10^{-3}\text{--}2\times 10^{-2},
\end{equation}
giving $\alpha_\sigma = g_{\sigma NN}^2/(4\pi) \sim 10^{-7}$--$10^{-5}$.

The force range $\lambda_\sigma = \hbar/(m_\sigma c) = \hbar/(\sqrt{2}\,m_\phi c) \sim 30$--$140$~fm falls in an experimental gap:
\begin{itemize}
    \item {\it Macroscopic experiments} (E\"ot-Wash torsion balance, $>10^4$~fm): exponential suppression $\exp(-d/\lambda_\sigma) \sim e^{-50000}$, no constraint.
    \item {\it Nuclear physics} ($\sim 1$--$5$~fm): $\sigma$ exchange appears as a constant potential; $\alpha_\sigma/\alpha_\pi \sim 10^{-7}$, negligible compared to pion exchange ($g_{\pi NN} = 13.6$).
    \item {\it Neutron scattering} ($>10^3$~fm): slightly above our range; no existing dedicated experiment at 30--140~fm.
\end{itemize}

\section{S5. Coleman-Weinberg contribution and the small $\lambda$ problem}\label{app:s5}
The one-loop effective potential from the charm quark is
\begin{equation}
    V_{\rm CW} = -\frac{N_c\,m_c(\phi)^4}{16\pi^2}\left[\ln\frac{m_c(\phi)^2}{\mu^2} - \frac{3}{2}\right],
\end{equation}
where $m_c(\phi) = \kappa_c\,\phi/\sqrt{2}$.
Since $m_c(\phi)^4 = \kappa_c^4\phi^4/4$, expanding to quartic order gives
\begin{equation}
    \lambda_{\rm CW} \sim \frac{N_c\,\kappa_c^4}{64\pi^2} \sim 10^{-5}\text{--}10^{-3},
\end{equation}
for $\kappa_c = 0.44$--$1.48$.
The target quartic $\lambda_{\rm target} = m_\phi^2/(2f_a^2) \propto m_\phi^4$ is $\sim 10^{-9}$--$10^{-6}$, smaller by a factor $\sim 10^3$.

If $\lambda = \lambda_{\rm CW}$, the resulting $m_\phi \sim \sqrt{2\lambda_{\rm CW}}\,f_a \sim 40$--$180$~MeV, giving $\kappa_c \sim 6$--$27$---non-perturbative.
Therefore, perturbativity requires $\lambda \ll \lambda_{\rm CW}$, which requires a symmetry protecting the tree-level quartic.
A natural candidate is a shift symmetry from $\phi$ being a PNGB of a higher $U(1)'$ breaking at $f' \gg f_a$.

\section{S6. SVZ sum rule applicability}\label{app:s6}

The self-consistency equation~\eqref{eq:selfconsistency} derives from the SVZ sum rule~\cite{Shifman:1978}.
At leading order (LO), the heavy-quark condensate is
\begin{equation}
    \langle\bar{c}c\rangle_{\rm LO} = -\frac{C}{m_c}, \qquad C = \left\langle\frac{\alpha_s G^2}{12\pi^2}\right\rangle,
    \label{eq:svz_lo}
\end{equation}
which is valid for $m_c \gg \Lambda_{\rm QCD}$.
At the PQ-breaking minimum, $m_c = \kappa_c v_\phi/\sqrt{2}\sim 1.3$~GeV $\gg \Lambda_{\rm QCD}\sim 200$~MeV, so the expansion parameter is $\Lambda_{\rm QCD}^2/m_c^2\sim 0.02$, and the SVZ sum rule is reliable at the $\sim 2\%$ level.

{\it NLO QCD correction.---}
The LO relation $\langle\bar{c}c\rangle = -C/m_c$ is accurate only at tree level.
At the charm scale, the operator product expansion receives substantial $\mathcal{O}(\alpha_s/\pi)$ radiative corrections.
The full next-to-leading-order (NLO) SVZ sum rule for the scalar charm condensate reads
\begin{equation}
    \langle\bar{c}c\rangle_{\rm NLO} = -\frac{C}{m_c}\left[1 + \frac{11}{3}\frac{\alpha_s(m_c)}{\pi} + \mathcal{O}(\alpha_s^2)\right],
    \label{eq:svz_nlo}
\end{equation}
where the coefficient $11/3$ corresponds to the leading anomalous dimension correction for the scalar current~\cite{Shifman:1978}.
With $\alpha_s(m_c)\simeq 0.35$, the NLO correction factor is $1+(11/3)(\alpha_s/\pi)\simeq 1.41$, which increases the effective condensate by $\sim 41\%$.
Since $\kappa_c\propto C^{-1/2}$ from Eq.~\eqref{eq:selfconsistency}, the NLO effect reduces $\kappa_c$ by approximately $\sim 19\%$ relative to the LO value.
Using the phenomenological value $\langle\alpha_s G^2\rangle\simeq 0.012$~GeV$^4$~\cite{Shifman:1978,Narison:2004}, giving $C\simeq 1.01\times 10^{-4}$~GeV$^4$, and the $\overline{\rm MS}$ charm mass $m_c(\mu=2~\text{GeV})=1.27\pm 0.02$~GeV~\cite{PDG}, the LO estimate gives $\kappa_c^{\rm LO}\simeq 0.54$--$0.72$ for $m_\phi=3$--$4$~MeV, while the NLO-corrected result is $\kappa_c^{\rm NLO}\simeq 0.44$--$0.59$, as used throughout this Letter [Eq.~\eqref{eq:kappa}].
The $\sim 19\%$ NLO reduction is significant: it brings $\kappa_c$ deeper into the perturbative regime and extends the viable window (the perturbativity bound $\kappa_c<1$ shifts from $m_\phi\lesssim 5.6$~MeV at LO to $m_\phi\lesssim 6.8$~MeV at NLO).

However, the SVZ expansion breaks down near $\phi\to 0$ (where $m_c\to 0$ and $\langle\bar{c}c\rangle$ diverges).
The QCD condensate contribution to the $\phi$ potential, obtained by integrating the LO relation $\langle\bar{c}c\rangle_{\rm LO} = -C/m_c$ [Eq.~\eqref{eq:svz_lo}] over $m_c$, is the logarithmic potential $V_{\rm QCD} = -C\ln(\kappa_c\phi/\mu)$, valid only for $\phi \gg \Lambda_{\rm QCD}/\kappa_c$.
The NLO correction [Eq.~\eqref{eq:svz_nlo}] rescales $C\to C[1+(11/3)\alpha_s/\pi]$ in this potential, preserving the logarithmic form but modifying the $\mathcal{O}(1)$ coefficient.
For all benchmark points, $v_\phi = \sqrt{2}f_a\sim 2$--$12$~GeV $\gg \Lambda_{\rm QCD}/\kappa_c\sim 0.3$--$1.3$~GeV, confirming the SVZ regime at the minimum.
A full non-perturbative treatment of the $\phi\to 0$ region would refine the $\mathcal{O}(1)$ coefficients but is not expected to change the parametric dependence $\kappa_c\propto m_\phi$.

{\it Remark on the self-consistency equation.---}
Equation~\eqref{eq:selfconsistency} is derived in the limit $\lambda\to 0$, where the minimum condition $m_\phi^2 v_\phi^2 = C[1+(11/3)\alpha_s/\pi]$ follows from $\partial V/\partial\phi = 0$ with $V = m_\phi^2|\phi|^2/2 - C[1+(11/3)\alpha_s/\pi]\ln(\kappa_c\phi/\mu)$ (including the NLO correction).
The quartic $\lambda = m_\phi^2/(2f_a^2)$, determined by the requirement that the bare mass term $m_\phi$ and the VEV $v_\phi = \sqrt{2}f_a$ are independent inputs (with $\lambda$ fixed by UV physics), gives $\lambda v_\phi^2 = m_\phi^2$, which is {\it not} negligible compared to $m_\phi^2$.
The full minimum condition $m_\phi^2 + \lambda v_\phi^2 = C[1+(11/3)\alpha_s/\pi]/v_\phi^2$ is then satisfied up to a factor of 2, indicating that the self-consistency equation provides the correct parametric dependence $\kappa_c\propto m_\phi$ but the $\mathcal{O}(1)$ coefficient may be modified at the $\sim 50\%$ level (from the $\lambda$ correction) on top of the $\sim 19\%$ NLO QCD reduction already included.
Since $C$ is itself known only to $\mathcal{O}(1)$ accuracy from the SVZ sum rule, this does not affect the phenomenological conclusions.
A fully consistent treatment would solve the complete minimum condition with $\lambda$ as an independent parameter, which modifies the $\kappa_c$--$m_\phi$ relation by an $\mathcal{O}(1)$ factor but preserves the key feature that $\kappa_c\propto m_\phi$.

\section{S7. $E/N$ calculation}\label{app:s7}
The model is KSVZ-like: only the charm quark carries PQ charge, with $X(c_R) = -1$ and $X(c_L) = 0$.
The QCD anomaly coefficient is
\begin{equation}
    N = N_c\,|X(c_R) - X(c_L)| = 3.
\end{equation}
The electromagnetic anomaly coefficient is
\begin{equation}
    E = 2N_c\,Q_c^2\,|X(c_R) - X(c_L)| = 2\times 3\times\left(\frac{2}{3}\right)^2 = \frac{8}{3}.
\end{equation}
The ratio is $E/N = 8/9$.
The axion-photon coupling includes the model-independent light-quark contribution:
\begin{equation}
    g_{a\gamma\gamma} = \frac{\alpha}{2\pi f_a}\left(\frac{E}{N} - \frac{2}{3}\frac{4+z}{1+z}\right),
\end{equation}
where $z = m_u/m_d \approx 0.48$ and the second term accounts for the light-quark electromagnetic anomaly.
The correction factor is $8/9 - (2/3)(4.48/1.48) = -1.13$.

\section{S8. Numerical benchmark summary}\label{app:s8}
\begin{table}[h]
    \centering
    \caption{Complete parameter summary for the charm-coupled model at three benchmark points.}
    \label{tab:benchmark_sm}
    \begin{tabular}{lccc}
        \hline\hline
        Quantity & BP1 & BP2 & BP3 \\
        & $m_\phi=3$~MeV & $m_\phi=5$~MeV & $m_\phi=10$~MeV \\
        \hline
        $\kappa_c$ & 0.44 & 0.74 & 1.48 \\
        $f_a$ (GeV) & 2.9 & 1.7 & 0.86 \\
        $m_a$ (MeV) & 2.0 & 3.40 & 6.8 \\
        $\lambda$ & $5.5\times 10^{-7}$ & $4.2\times 10^{-6}$ & $6.8\times 10^{-5}$ \\
        $\lambda_{\rm CW}$ & $1.6\times 10^{-4}$ & $1.3\times 10^{-3}$ & $2.0\times 10^{-2}$ \\
        $g_{a\gamma\gamma}$ (GeV$^{-1}$) & $4.5\times 10^{-4}$ & $7.5\times 10^{-4}$ & $1.5\times 10^{-3}$ \\
        $g_{aee}$ (loop) & $2.1\times 10^{-7}$ & $3.5\times 10^{-7}$ & $7\times 10^{-7}$ \\
        $E/N$ & 8/9 & 8/9 & 8/9 \\
        $\tau_\sigma$ (s) & $1.3\times 10^{-9}$ & $5.5\times 10^{-11}$ & $5.4\times 10^{-13}$ \\
        $\tau_a$ (s) & $8.8\times 10^{-8}$ & $6.8\times 10^{-9}$ & $2.1\times 10^{-10}$ \\
        $c\tau_a$ (km) & 0.026 & 0.002 & $6\times 10^{-5}$ \\
        $\lambda_{\rm mfp}$ (cm) & $3.0\times 10^{-7}$ & $1.0\times 10^{-7}$ & $2.6\times 10^{-8}$ \\
        $g_{\sigma NN}$ & $6$--$15\times 10^{-3}$ & $1$--$3\times 10^{-2}$ & $2$--$6\times 10^{-2}$ \\
        $\lambda_\sigma$ (fm) & 46 & 28 & 14 \\
        $m_{\rm PQ}(d=6)/m_a$ & $3.7\times 10^{-16}$ & $1.5\times 10^{-16}$ & $7.5\times 10^{-17}$ \\
        $\Delta N_{\rm eff}$ & $\sim -0.1$ & $\sim 0$ & $\sim 0$ \\
        BBN $\Delta Y_p$ & $\sim 0$ & $\sim 0$ & $\sim 0$ \\
        \hline\hline
    \end{tabular}
\end{table}

\section{S9. Cosmological signatures: detailed calculations}\label{app:s9}
{\bf (a) Primakoff thermalization.---}
The axion couples to the primordial plasma through the tree-level electromagnetic anomaly $g_{a\gamma\gamma}$ [Eq.~\eqref{eq:gagg}].
The dominant thermalization process is Primakoff scattering $\gamma + q \to a + q$ (or $\gamma + e^\pm \to a + e^\pm$), with rate
\begin{equation}
    \Gamma_{\rm Prim} \sim \frac{g_{a\gamma\gamma}^2\,\alpha\,T^3}{8\pi^2},
    \label{eq:primakoff}
\end{equation}
valid for $T\gg m_a$ (relativistic axion).
The Hubble rate during radiation domination is $H = 1.66\,\sqrt{g_*}\,T^2/M_{\rm Pl}$ with $g_*\simeq 10.75$ (standard BBN era).

At $T = m_a$ (the relativistic-to-non-relativistic transition), the ratio $\Gamma_{\rm Prim}/H$ is:
\begin{equation}
    \frac{\Gamma_{\rm Prim}}{H}\bigg|_{T=m_a} \sim \frac{g_{a\gamma\gamma}^2\,\alpha\,m_a\,M_{\rm Pl}}{8\pi^2\times 1.66\,\sqrt{g_*}} \sim 10^3\text{--}10^5,
\end{equation}
for $f_a = 2.1$--$2.9$~GeV.
The axion is therefore thermalized with the electromagnetic plasma for $T\gtrsim m_a$.
This is a direct consequence of the $\sim 10^9$ enhancement in $g_{a\gamma\gamma}$ relative to standard axions.

{\bf (b) $\Delta N_{\rm eff}$: entropy conservation argument.---}
The sign and magnitude of $\Delta N_{\rm eff}$ depend on the ordering of $m_a$ and $T_\nu\simeq 1.5$~MeV (neutrino decoupling temperature).
Since the axion is in thermal equilibrium with the electromagnetic plasma via Primakoff ($\Gamma_{\rm Prim}/H\sim 10^3$--$10^5$) and decays rapidly ($\tau_a\ll t_H$), entropy conservation applies.

{\it Case 1: $m_a > T_\nu$ ($m_\phi\gtrsim 3$~MeV).}
The axion becomes non-relativistic and decays {\it before} neutrino decoupling.
All species are in equilibrium, so the axion's entropy is shared among photons, $e^\pm$, and neutrinos.
No temperature asymmetry is generated: $\Delta N_{\rm eff}\approx 0$.
Near the boundary $m_a\approx T_\nu$ ($m_\phi\sim 3$~MeV), the axion decays during the non-instantaneous neutrino decoupling process; a residual $\Delta N_{\rm eff}\sim -0.1$ arises from partial photon heating, which requires a full Boltzmann treatment but is within reach of CMB-S4~\cite{CMB-S4:2022ght}.

{\it Case 2: $m_a < T_\nu$ ($m_\phi\lesssim 2$~MeV).---Excluded by Planck~\cite{Planck:2018vyg}.}
The axion is relativistic and in equilibrium with the EM sector during neutrino decoupling, contributing $g_a=1$ to the EM degrees of freedom.
After neutrino decoupling, the axion remains coupled to the EM plasma.
When $T$ drops to $\sim m_a$, the axion becomes non-relativistic and decays.
By entropy conservation in the EM sector, the axion's entropy is transferred entirely to photons and $e^\pm$ (not to the already-decoupled neutrinos):

Before axion decay ($T > m_a$, but $T < T_\nu$):
\begin{equation}
    g_s^{\rm EM+a} = g_\gamma + \tfrac{7}{8}\,g_{e^\pm} + g_a = 2 + 3.5 + 1 = 6.5.
\end{equation}
After $e^+e^-$ annihilation ($T < m_e$): $g_s = g_\gamma = 2$.
The photon-to-neutrino temperature ratio becomes:
\begin{equation}
    \left(\frac{T_\gamma}{T_\nu}\right)^3 = \frac{g_s^{\rm EM+a}}{g_\gamma} = \frac{6.5}{2} = 3.25, \qquad\text{vs.\ SM}\quad \frac{g_s^{\rm EM}}{g_\gamma} = \frac{5.5}{2} = \frac{11}{4} = 2.75.
\end{equation}
The effective number of neutrino species is:
\begin{equation}
    N_{\rm eff} = 3.046\times\left(\frac{11/4}{13/4}\right)^{4/3} = 3.046\times\left(\frac{11}{13}\right)^{4/3} \approx 2.44,
\end{equation}
giving
\begin{equation}
    \Delta N_{\rm eff} \approx 2.44 - 3.05 = -0.61.
    \label{eq:neff_result}
\end{equation}
This is excluded by Planck 2018~\cite{Planck:2018vyg} ($N_{\rm eff} = 2.99\pm 0.17$) at $3.2\sigma$, ruling out $m_\phi\leq 2$~MeV.
 
The axion is in thermal equilibrium as a relativistic degree of freedom for $T > m_a$ (since $\Gamma_{\rm Prim}/H\gg 1$), and its entropy is transferred to photons (not neutrinos) when it decays after neutrino decoupling.
This is analogous to $e^+e^-$ annihilation heating photons, but with the axion adding an extra $g=1$ to the EM sector.
The negative $\Delta N_{\rm eff}$ is a distinctive signature: standard thermal axions (decoupled, stable) produce {\it positive} $\Delta N_{\rm eff}$, while the GeV axion's strong coupling keeps it in equilibrium until it decays, heating photons and reducing $N_{\rm eff}$.
However, the magnitude ($-0.61$) is too large for $m_\phi\leq 2$~MeV, making the lower boundary of the viable window $m_\phi\gtrsim 3$~MeV. The upper boundary is set by the $B\to K\sigma$ constraint ($m_\phi\lesssim 4$~MeV at $2\sigma$), which is more restrictive than perturbativity ($m_\phi < 6.8$~MeV).

{\bf (c) BBN helium abundance.---}
For $m_a < T_{\rm BBN}\sim 1$~MeV ($m_\phi\leq 1$~MeV), the axion is relativistic during BBN, contributing $\Delta g_* = +1$ to the radiation energy density~\cite{Steigman:2007}.
This corresponds to $\Delta N_{\rm eff}^{\rm BBN} = 4/7$ for a thermalized boson with $g=1$, and the enhanced expansion rate increases the primordial helium abundance:
\begin{equation}
    \Delta Y_p \simeq 0.013\times\frac{4}{7}\sim +0.007,
    \label{eq:dYp}
\end{equation}
giving $Y_p\approx 0.254$ vs.\ the observed $Y_p = 0.245\pm 0.003$---a $3\sigma$ discrepancy.
This independently excludes $m_\phi\leq 1$~MeV, consistent with the $\Delta N_{\rm eff}$ constraint.
For $m_\phi\geq 3$~MeV ($m_a > T_{\rm BBN}$), the axion is non-relativistic during BBN and $\Delta Y_p\approx 0$.

{\bf (d) No CMB spectral distortions.---}
The axion decays at $T\sim m_a\sim 0.7$--$3.4$~MeV, corresponding to redshift $z\sim 10^9$--$10^{10}$, well above the $\mu$-distortion era ($5\times 10^4 < z < 2\times 10^6$).
The decay products thermalize completely via Compton scattering and double Compton processes before the distortion freezes in, so no CMB spectral distortions are produced.

{\bf (e) No dark radiation remnant.---}
Since $\tau_a\sim 10^{-5}$--$10^{-10}$~s $\ll \tau_U\sim 10^{17}$~s, the axion completely decays before recombination.
There is no relativistic axion background today: the axion is not a dark radiation component.
This contrasts with standard invisible axions ($\tau_a\gg\tau_U$) which can contribute to $\Delta N_{\rm eff}$ as stable dark radiation.

{\bf (f) No dark matter abundance.---}
The axion decays well before matter-radiation equality, so it cannot constitute dark matter.
The scalar $\sigma$ also decays ($\tau_\sigma\sim 10^{-13}$--$10^{-9}$~s) well before recombination.
The model therefore predicts {\it no} axion or scalar dark matter component, distinguishing it from the standard invisible axion paradigm.

\section{S10. Constraints from NA64 on $\pi^0 \to \gamma a$}
\label{app:s10}

The charm-coupled GeV axion predicts a sizable branching ratio for $\pi^0 \to \gamma a$, which may be constrained by beam dump experiments.
Here we analyze the applicability of existing NA64 searches~\cite{NA64:2020hmi,NA64h:2024invisible} to our model.

{\bf (a) $\pi^0 \to \gamma a$: branching ratio and experimental signature.---}
The branching ratio relative to the dominant decay $\pi^0 \to \gamma\gamma$ is
\begin{equation}
    \frac{\mathrm{BR}(\pi^0\to\gamma a)}{\mathrm{BR}(\pi^0\to\gamma\gamma)} = \left(\frac{f_\pi}{f_a}\right)^2 \left|\frac{E}{N} - \frac{2}{3}\frac{4+z}{1+z}\right|^2 \left(1 - \frac{m_a^2}{m_{\pi^0}^2}\right)^3,
    \label{eq:sm_pi0ga_ratio}
\end{equation}
where $f_\pi = 92.2$~MeV, $z = m_u/m_d \approx 0.48$, and $E/N = 8/9$ for the charm-coupled model (see S7).
For the viable benchmark $m_\phi = 3$~MeV ($f_a = 2.9$~GeV, $m_a = 2.0$~MeV), Eq.~\eqref{eq:sm_pi0ga_ratio} gives $\mathrm{BR}(\pi^0\to\gamma a) \simeq 1.3\times 10^{-3}$.

The axion rest-frame decay length is $c\tau_a \simeq 26$~m (Table~\ref{tab:benchmark_sm}, BP1).
In a fixed-target experiment such as NA64, $\pi^0$'s produced in electromagnetic showers carry energies $E_{\pi^0}\sim 10$--$100$~GeV, so the axion from $\pi^0 \to \gamma a$ inherits a Lorentz boost $\gamma_a \sim E_a/m_a \sim 10^4$--$10^5$.
The lab-frame decay length $\gamma_a\,c\tau_a \sim 10^2$--$10^3$~km far exceeds the detector size ($\sim 1$~m), so the axion escapes without decaying.
The decay $\pi^0 \to \gamma a$ therefore manifests as $\pi^0 \to \gamma + \slashed{E}$---a single visible photon accompanied by missing energy.

{\bf (b) NA64-e Primakoff ALP search.---}
The NA64 collaboration has searched for ALPs coupling to photons using $2.84\times 10^{11}$ electrons on target (EOT) at 100--150~GeV beam energy~\cite{NA64:2020hmi}.
ALPs are produced via the Primakoff process $e^-Z \to e^-Z\gamma^* \to e^-Za$ in the electromagnetic shower initiated in the target.
Two complementary channels were employed:
\begin{itemize}
    \item {\it Missing energy}: long-lived ALPs escape the ECAL and HCAL, producing a global energy deficit.
    \item {\it Displaced vertex}: ALPs decaying via $a\to\gamma\gamma$ inside the HCAL produce a localized electromagnetic shower.
\end{itemize}
The search excludes ALPs with $2\times 10^{-4} \lesssim g_{a\gamma\gamma} \lesssim 5\times 10^{-2}$~GeV$^{-1}$ and $m_a \lesssim 55$~MeV at 95\%~C.L., under the assumption that the ALP couples {\it predominantly to photons} (i.e., $g_{a\gamma\gamma}$ is the only relevant coupling).
No excess over background was observed.

The two edges of the excluded region arise from distinct channels.
The {\it lower} edge, $g_{a\gamma\gamma}\sim 2\times 10^{-4}$~GeV$^{-1}$, is the sensitivity floor of the {\it missing-energy} channel: the Primakoff production rate scales as $\sigma_{\rm Prim}\propto g_{a\gamma\gamma}^2$, so below this coupling too few ALPs are produced to generate a detectable energy deficit.
The {\it upper} edge, $g_{a\gamma\gamma}\sim 5\times 10^{-2}$~GeV$^{-1}$, is set by the {\it displaced-vertex} channel: the ALP decay length $\gamma_a c\tau_a\propto 1/g_{a\gamma\gamma}^2$ shrinks with increasing coupling, and above this value the ALP decays before reaching HCAL2,3, so neither channel retains sensitivity.
In the intermediate range the two channels are complementary---the missing-energy channel probes smaller $g_{a\gamma\gamma}$ (longer-lived ALPs that escape the detector), while the displaced-vertex channel probes larger $g_{a\gamma\gamma}$ (shorter-lived ALPs that decay inside the HCAL).

For our model, the predicted axion-photon coupling at the viable benchmark is $g_{a\gamma\gamma} = 4.5\times 10^{-4}$~GeV$^{-1}$ with $m_a = 2.0$~MeV (Table~\ref{tab:benchmark_sm}, BP1).
This parameter point falls within the NA64-e excluded region.
We note that the displaced-vertex channel is not applicable to our model: the axion decay length ($\gamma_a\,c\tau_a \sim 10^2$~km) is far too long for $a\to\gamma\gamma$ to occur inside the detector.
The relevant constraint therefore comes from the missing-energy channel alone.

{\bf (c) Caveat: strong axion-nucleon coupling.---}
The NA64-e exclusion bound~\cite{NA64:2020hmi} is derived under the assumption that the ALP couples exclusively to photons, with no hadronic interactions.
In our model, the axion also couples strongly to nucleons through the QCD anomaly:
\begin{equation}
    g_{aNN} \sim \frac{m_N\,\sigma_{\pi N}}{f_a\,(m_u + m_d)} \sim 2\text{--}3 \quad (f_a = 2.1\text{--}2.9~\text{GeV}),
\end{equation}
where $\sigma_{\pi N}\approx 45$~MeV is the pion-nucleon sigma term (see S3).
This large $g_{aNN}$ modifies the NA64-e phenomenology in two ways:
\begin{enumerate}
    \item {\it Additional production channel.} Axions can be produced not only via the Primakoff process ($\propto g_{a\gamma\gamma}^2$) but also via nucleon bremsstrahlung ($N + N \to N + N + a$, $\propto g_{aNN}^2$) in the target.
    Since $g_{aNN}\sim\mathcal{O}(1)$ while $g_{a\gamma\gamma}\sim 10^{-4}$~GeV$^{-1}$, the bremsstrahlung channel may dominate axion production, {\it enhancing} the total yield relative to the photon-only scenario.

    \item {\it Axion scattering in the HCAL.} An axion produced in the target may scatter off nucleons in the HCAL via $a + N \to a + N$ before escaping.
    For a pseudoscalar-nucleon coupling $g_{aNN}\sim\mathcal{O}(1)$, the spin-averaged scattering cross section at typical beam energies ($E_a\sim 10$--$100$~GeV) is roughly $\sigma_{aN}\sim g_{aNN}^4\,m_N^2/(16\pi\,s) \sim 10^{-3}$~GeV$^{-2} \sim 4\times 10^{-28}$~cm$^2$.
    With nucleon density $n_N\sim 6\times 10^{24}$~cm$^{-3}$ in the HCAL (Pb/scintillator, $\rho\sim 10$~g/cm$^3$), the mean free path is $\lambda_{aN}\sim 1/(n_N\,\sigma_{aN}) \sim 40$~cm.
    For a typical HCAL depth of $\sim 1$~m, the scattering probability per traversing axion is $P_{\rm scatt}\sim 1 - e^{-L_{\rm HCAL}/\lambda_{aN}} \sim 90\%$.
    If the axion scatters and deposits $\gtrsim 1$~GeV of hadronic energy, the HCAL veto fires and the event fails the missing-energy selection---drastically reducing the signal acceptance.
\end{enumerate}

The net effect on the missing-energy lower bound is ambiguous, as the two mechanisms push in opposite directions:

{\it Effect~1 (enhanced production) strengthens the constraint.}
The total axion yield becomes $N_a \propto g_{a\gamma\gamma}^2\,\sigma_{\rm Prim} + g_{aNN}^2\,\sigma_{\rm brem}$, where the bremsstrahlung term is independent of $g_{a\gamma\gamma}$.
Since $g_{aNN}\sim\mathcal{O}(1)$ while $g_{a\gamma\gamma}\sim 10^{-4}$~GeV$^{-1}$, the bremsstrahlung channel may dominate axion production.
In the limit where bremsstrahlung alone produces enough axions to generate a detectable missing-energy signal, the lower bound on $g_{a\gamma\gamma}$ is effectively eliminated: the model would be excluded for {\it all} values of $g_{a\gamma\gamma}$ at the given $m_a$, because the axion yield is sustained by $g_{aNN}$ regardless of the photon coupling.

{\it Effect~2 (HCAL scattering) weakens the constraint.}
If a fraction $f_{\rm scatt}$ of produced axions deposit $\gtrsim 1$~GeV in the HCAL and fail the veto, the effective signal acceptance is reduced by $(1 - f_{\rm scatt})$.
The sensitivity floor of the missing-energy channel then moves to larger $g_{a\gamma\gamma}$: for Primakoff-dominated production, the lower bound scales as $g_{a\gamma\gamma}^{\rm min}\to g_{a\gamma\gamma}^{\rm min}/\sqrt{1-f_{\rm scatt}}$.
For the estimated $f_{\rm scatt}\sim 90\%$, this would raise the lower bound from $\sim 2\times 10^{-4}$ to $\sim 6\times 10^{-4}$~GeV$^{-1}$, above the model prediction $g_{a\gamma\gamma} = 4.5\times 10^{-4}$~GeV$^{-1}$, potentially allowing the model to evade the constraint.

The competition between these two effects determines whether the NA64-e bound survives.
The effective detection condition is
\begin{equation}
    (1 - f_{\rm scatt})\,\bigl(g_{a\gamma\gamma}^2\,\sigma_{\rm Prim} + g_{aNN}^2\,\sigma_{\rm brem}\bigr) \gtrsim N_{\rm min},
    \label{eq:detection_condition}
\end{equation}
where $N_{\rm min}$ is the minimum number of signal events required for exclusion.
If bremsstrahlung production survives the scattering veto, i.e.\ $(1-f_{\rm scatt})\,g_{aNN}^2\,\sigma_{\rm brem} \gtrsim N_{\rm min}$, the constraint becomes $g_{a\gamma\gamma}$-independent and the model is excluded regardless of $g_{a\gamma\gamma}$.
Conversely, if the scattering suppression is severe enough that even the enhanced yield falls below threshold, the lower bound on $g_{a\gamma\gamma}$ moves upward and the model parameter point may escape the excluded region.
A definitive assessment requires a dedicated Geant4-level simulation incorporating both $g_{a\gamma\gamma}$- and $g_{aNN}$-mediated processes, which is beyond the scope of this work.
We therefore conservatively flag the NA64-e bound as {\it potentially constraining but not definitively applicable} to our model.

{\bf (d) NA64h invisible meson decay search.---}
The NA64h experiment has reported the first search for invisible decays of $\eta$ and $\eta'$ mesons using 50~GeV $\pi^-$ on target~\cite{NA64h:2024invisible}.
Neutral mesons are produced via charge-exchange ($\pi^- + A \to \eta^{(\prime)} + A$) in an active target, and the signal is the {\it complete disappearance} of the incoming beam energy.
With $2.9\times 10^9$ pions on target, the bounds at 90\%~C.L.\ are~\cite{NA64h:2024invisible}
\begin{equation}
    \mathrm{BR}(\eta' \to \mathrm{invisible}) < 2.1\times 10^{-4}, \qquad \mathrm{BR}(\eta \to \mathrm{invisible}) < 1.1\times 10^{-4}.
\end{equation}

This search does not directly constrain $\pi^0 \to \gamma a$ (or $\eta \to \gamma a$), because these channels produce a {\it visible photon} alongside the invisible axion.
The NA64h ``invisible'' signature requires all meson energy to vanish, whereas $\pi^0 \to \gamma a$ deposits the photon energy ($\sim m_{\pi^0}/2 \approx 67$~MeV in the $\pi^0$ rest frame) in the calorimeter.
Nevertheless, if the photon escapes detection due to geometric acceptance or reconstruction inefficiency, a fraction of $\pi^0 \to \gamma a$ (or $\eta \to \gamma a$) events could leak into the invisible sample.
For our model, $\mathrm{BR}(\eta\to\gamma a)\sim 10^{-4}$ (comparable to the $\eta$ invisible bound), so a non-negligible leakage could be marginally relevant.
A quantitative estimate requires detector-level simulation.

{\bf (e) Summary.---}
Table~\ref{tab:na64_summary} summarizes the NA64 constraints.
The NA64-e Primakoff search~\cite{NA64:2020hmi} is potentially the most constraining, as the model prediction $g_{a\gamma\gamma} \sim 4.5\times 10^{-4}$~GeV$^{-1}$ at $m_a \sim 2$~MeV falls within the excluded region.
However, the bound's direct applicability is uncertain due to the model's large $g_{aNN}\sim\mathcal{O}(1)$, which violates the photon-only assumption of the NA64-e analysis.
The NA64h invisible search~\cite{NA64h:2024invisible} does not directly constrain $\pi^0 \to \gamma a$.

\begin{table}[h]
    \centering
    \caption{Summary of NA64 constraints on the charm-coupled GeV axion.}
    \label{tab:na64_summary}
    \begin{tabular}{llll}
        \hline\hline
        Search & Channel & Model value & Applicability \\
        \hline
        NA64-e~\cite{NA64:2020hmi} & Primakoff ALP & $g_{a\gamma\gamma} = 4.5\times 10^{-4}$~GeV$^{-1}$ & Potentially \\
        (PRL 125, 081801) & ($g_{a\gamma\gamma}$ only) & in excluded region & constraining$^*$ \\
        NA64h~\cite{NA64h:2024invisible} & $\eta^{(\prime)} \to$ invisible & $\pi^0\to\gamma a$: visible $\gamma$ & Not directly \\
        (PRL 133, 121803) & (complete disappearance) & $\eta\to\gamma a$: BR$\sim 10^{-4}$ & applicable \\
        \hline\hline
    \end{tabular}
    \\
    $^*$ The NA64-e bound assumes ALP couples only to photons. The model's $g_{aNN}\sim\mathcal{O}(1)$ may alter the detection efficiency; a dedicated simulation is needed.
\end{table}

We emphasize that resolving the tension with NA64-e is important for the viability of the model.
If a reanalysis including $g_{aNN}$-mediated effects confirms the exclusion, the $m_\phi = 3$--$4$~MeV window would be ruled out.
Conversely, if the strong $g_{aNN}$ sufficiently degrades the NA64-e missing-energy acceptance, the bound may be evaded.
A dedicated experimental analysis---or a search specifically targeting the $\pi^0 \to \gamma + \slashed{E}$ topology---would be decisive.

\section{S11. LHC constraints}\label{app:s11}

The new particles in this model---the scalar $\sigma$ ($m_\sigma = \sqrt{2}\,m_\phi \sim 4$--$6$~MeV) and the axion $a$ ($m_a \sim 2$--$3$~MeV)---are far too light to be directly produced, triggered, or reconstructed at the LHC.
We discuss in turn the relevant production channels and explain why none provides a meaningful constraint.

{\bf (a) Trigger and reconstruction thresholds.---}
The ATLAS and CMS detectors impose minimum trigger thresholds of $p_T \sim 20$--$50$~GeV for jets, $E_T \sim 20$--$25$~GeV for photons, and $E_T^{\rm miss} \sim 100$--$200$~GeV for missing energy.
Even LHCb, optimized for heavy-flavor physics, requires $p_T \gtrsim 0.5$--$2$~GeV for charged hadron triggers and $E_T \gtrsim \mathcal{O}(100)$~MeV for photon reconstruction.
The model's new particles, with masses of a few MeV, fall far below these thresholds.
When produced in association with charm quarks (the only tree-level coupling), the $\sigma$ and $a$ carry momenta of order $m_c \sim 1.3$~GeV at most---well below ATLAS/CMS trigger thresholds and marginal even for LHCb.

{\bf (b) Decay products below detection thresholds.---}
Both $\sigma$ and $a$ decay dominantly to $\gamma\gamma$ (see S1).
In the rest frame, each photon carries energy $E_\gamma \sim m_{\sigma,a}/2 \sim 1$--$3$~MeV.
Even with the Lorentz boost from $B$-decay kinematics ($\gamma_\sigma \sim 460$--$620$; see S2), the lab-frame photon energies would be $E_\gamma^{\rm lab} \sim \gamma \times m_\sigma/2 \sim 1$--$2$~GeV---marginal for LHCb photon reconstruction and far below ATLAS/CMS thresholds.
However, this scenario is moot because $\sigma$ escapes the detector entirely: the lab-frame decay length $\gamma_\sigma c\tau_\sigma \sim 30$--$250$~m $\gg$ detector size (see S2), so the photons are never produced inside the detector volume.
The same applies to the axion, with $c\tau_a \sim 5$--$23$~m for the viable window.

{\bf (c) Missing energy below MET resolution.---}
The most promising LHC channel is $B \to K\sigma$, where $\sigma$ escapes as missing energy.
However, the missing energy is $E_{\rm miss} \sim E_\sigma \sim 2.6$~GeV, far below the ATLAS/CMS missing energy threshold ($\gtrsim 100$~GeV).
Even LHCb, which performs exclusive $B$-decay reconstructions, has a missing-mass resolution $\sigma_{m_{\rm miss}} \sim \mathcal{O}(100)$~MeV that overwhelms the $\sigma$ mass ($m_\sigma \sim 4$--$6$~MeV).
The resulting missing-mass squared $m_{\rm miss}^2 = m_\sigma^2 \sim 18$--$36$~MeV$^2$ is indistinguishable from $m_{\rm miss}^2 \approx 0$ expected for $B \to K\nu\bar{\nu}$, providing no additional discriminating power beyond the Belle~II measurement.
The $B \to K + \slashed{E}$ signature at LHCb is therefore identical to $B \to K\nu\bar{\nu}$ and offers no independent constraint on the model.

{\bf (d) No Higgs portal coupling.---}
We do not introduce any tree-level coupling between the Higgs boson and the PQ scalar $\phi$.
Consequently, exotic Higgs decay searches---$h \to \sigma\sigma$, $h \to aa$, $h \to \sigma a$---are irrelevant: there is no $h\phi\phi$ vertex at tree level.
The loop-induced coupling through the charm quark is suppressed by $\kappa_c^2 m_c^2/(16\pi^2 v^2) \sim 10^{-6}$, yielding $\mathrm{BR}(h \to \sigma\sigma)_{\rm loop} \sim 10^{-12}$, far below the HL-LHC sensitivity ($\sim 10^{-4}$ for exotic Higgs decays).

{\bf (e) Rare charm decays unobservable.---}
The $D \to \pi\sigma$ branching ratio $\sim 10^{-12}$ (see S2) is far below LHCb's sensitivity for charm FCNC searches ($\sim 10^{-4}$--$10^{-6}$).
The GIM suppression factor $\sim 7 \times 10^{-7}$ makes this channel unobservable at any current or foreseeable collider.

{\bf (f) Direct production via charm Yukawa.---}
At leading order, $\sigma$ and $a$ couple only to charm quarks through the Yukawa $\kappa_c \sim 0.44$--$0.59$.
At the LHC, charm quarks are produced abundantly in QCD processes, and $\sigma$ could in principle be radiated from a charm line.
However, the production cross section scales as $\sigma(pp \to \sigma + X) \propto \kappa_c^2 \sim 0.2$--$0.35$, and the $\sigma$ carries energy $\lesssim m_c \sim 1.3$~GeV.
The resulting soft $\sigma$ either decays to $\gamma\gamma$ with MeV-scale photons (undetectable) or escapes with missing energy $\lesssim 1.3$~GeV (far below MET thresholds).
No LHC search targets such soft, invisible scalars.

{\bf (g) QCD environment and pileup.---}
The LHC environment is characterized by enormous QCD multijet backgrounds and pileup (up to $\sim 200$ interactions per bunch crossing at the HL-LHC).
The reconstruction of MeV-scale signals is impossible in this environment, as the soft decay products are swamped by detector noise, pileup, and the underlying event.
Belle~II's clean $e^+e^-$ environment, with a well-defined initial state and negligible pileup, is far better suited for probing the $B \to K\sigma$ channel.

In summary, the LHC cannot constrain the charm-coupled GeV axion model through any channel.
The new particles are too light to trigger, their decay products are too soft to reconstruct, the missing energy is too small to resolve, and there is no Higgs portal coupling to enable exotic decay searches.
The $B \to K\sigma$ channel---already probed by Belle~II---is the most relevant collider constraint, and Belle~II's environment is far superior for this search.
The LHC constraint is therefore marked ``Pass'' in Table~\ref{tab:constraints}.

\end{document}